# OUTLIERS EMPHASIS ON CLUSTER ANALYSIS

The use of squared Euclidean distance and fuzzy clustering to detect outliers in a dataset

Gianluca Rosso[1]

ABSTRACT

Outlier is the term that indicates in statistics an anomalous observation, aberrant, clearly distant from others collected observations. The outliers are the subject to animated discussions in various contexts with regard to be or not to be considered in the average evaluations. Outliers can become a precious source of information, on condition that be able to accurately identify the presence in the reference datasets. The need to identify the presence of clustered outliers in a dataset not previously treated could argue for a fuzzy clustering, emphasized by using the quadratic Euclidean distance as similarity measure. For interesting and useful results, it should be inclined a possibilistic clustering approach, where the term "possibilistic" means, always in mathematical rigor, a component of interpretation of values that point out anomalous cases. The crisp method does not allow it, the fuzzy method introduce it, the possibilistic one use it.

This is a very simple paper with divulgative purposes, addressed especially to students, but not only.

KEYWORDS: cluster analysis, crisp clustering, fuzzy clustering, squared Euclidean distance, possibilistic clustering, outliers.

---

[1] *Gianluca Rosso is Graduate Statistician (GradStat) at RSS The Royal Statistical Society in London, and Correspondent Researcher (Full Associate) at SIS Italian Statistical Society in Rome (gianluca.rosso@sis-statistica.org).*

1. INTRODUCTION.

Outlier is the term that indicates in statistics an anomalous observation, aberrant, clearly distant from others collected observations.

The descriptive statistics from datasets containing outliers, may provide highly misleading keys to understanding phenomenons. Understand its origin is not always easy, but it is definitely necessary. But it is even more important to know what to do, how to treat them and how to identify them uniquely.

The outliers are the subject to animated discussions in various contexts with regard to be or not to be considered in the average evaluations. For example in the business sector, in which it is not uncommon that the budget achievement is carried out by isolated, and perhaps not repeatable, large-scale contracts. Or in the insurance sector, and particularly in the claim sector in which may exist amounts so-called catastrophic, rare to happen but causing highly distorted statistics. Now think at the context of healthness and epidemiology: a group of cancer patients contains one or two people who respond to treatment in a markedly different and positive way: they could raise the average errors of assessment in inducing doses of therapy, to the detriment of other patients while belonging to the group but that respond to treatment in a more modest way.

In the last two years the study of outliers was directed with great determination to fraud cases searching, especially financial, and to internet intrusion, once and for all acquiring authoritativeness in the field of risk analysis. The study of these cases is able to highlight abnormal behavior characteristics that are not always random, but often hide fraudulent and volunteers behavior. Many internal investigations for companies, but also of the police or tax authorities, are based on the search for important quantitative abnormalities in the enormous mass of available data.

From the examples above is possibile to understand how outliers can become a precious source of information, on condition that be able to accurately identify the presence in the reference datasets. The same outliers become a touchstone in extreme situations for quality control: if the point of breaking of a ceramic tile of the Shuttle had become a simple value among many, or worse, had been put outside from the database to avoid distortions, almost certainly no Shuttle would never return to earth.

In cluster analysis, especially in the presence of numerically relevant datasets, it can be difficult to identify outlier, especially if you have not done a careful prior examination of the dataset.

The same choise of the clustering method should be carefully considered. It is not always required to use the most hard method (crisp clustering) for the group analysis, or rather in some cases may not detect some particularities of the dataset. In many cases, the fuzzy clustering can be an effective and perhaps better alternative. But, as we shall see later, the need to identify the presence of clustered outliers in a dataset not previously treated could argue for a fuzzy clustering, emphasized by using the quadratic Euclidean distance as similarity measure. For interesting and useful results, it should be inclined a possibilistic clustering approach, where the term "possibilistic" means, always in mathematical rigor, a component of interpretation of values that point out anomalous cases. The crisp method does not allow it, the fuzzy method introduce it, the possibilistic one use it.

The statistical methods related to possibilistic clustering are introduced many times ago (in 1997 by N. Pal, K. Pal, J. Keller, J. Bezdek), and have a difficulty level that is not much suitable for practical use. From the Possibilistic Theory will be maintained basic assumption regarding cancellation of the membership cluster constraints; therefore here will be proposed a theoretical case and a simple method to get a reliable result by using a simple spreadsheet.

1. CRISP CLUSTERING.

We use a tipical study dataset in traditional clustering, which is a symmetric dataset.

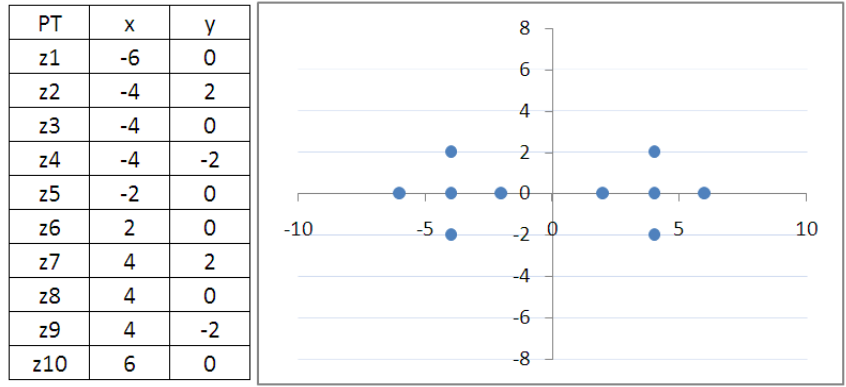

| PT  | x  | y  |
|-----|----|----|
| z1  | -6 | 0  |
| z2  | -4 | 2  |
| z3  | -4 | 0  |
| z4  | -4 | -2 |
| z5  | -2 | 0  |
| z6  | 2  | 0  |
| z7  | 4  | 2  |
| z8  | 4  | 0  |
| z9  | 4  | -2 |
| z10 | 6  | 0  |

*Fig.1*

In crisp clustering, data will be organized into groups with maximum internal homogeneity and maximum external heterogeneity. The membership is done by calculating and minimizing the distance from an average point, called centroid. To get the final result probably is needed a new displacement of the centroid (reiteration) for several times, and recalculation of the Euclidean distances of points, until the recalculation does not supply more changes in the cluster and thus does indicate that it has reached the position of balance status, and conditions of the crisp method are verified, ie

$$\mu_{ik} \in \{0,1\}, \quad 1 \leq i \leq c, \quad 1 \leq k \leq N \qquad (2.1a)$$

$$\sum_{i=1}^{c} \mu_{ik} = 1, \quad 1 \leq k \leq N \qquad (2.1b)$$

$$0 < \sum_{k=1}^{N} \mu_{ik} < N, \quad 1 \leq i \leq c \qquad (2.1c)$$

In crisp clustering data is clearly attributed (rightly or wrongly) to a cluster. Imagine that you want to attribute the data in our possession within two clusters.

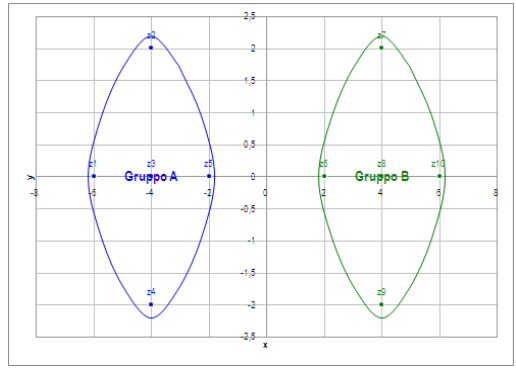

*Fig. 2*

The centroids are located respectively in the overlapping elements of Z3 and Z8.

The table of allocation would be as follows.

$$U = \begin{bmatrix} 1 & 1 & 1 & 1 & 1 & 0 & 0 & 0 & 0 & 0 \\ 0 & 0 & 0 & 0 & 0 & 1 & 1 & 1 & 1 & 1 \end{bmatrix}$$

*tab.1*

The case becomes complicated when we insert an additional data that arises in exactly symmetrical and equidistant position from the two clusters calculated.

This is clearly an extreme case, probably not much related to reality, but this explain the difficulty of decision that has to take the analyst. Following the rules described in (2.1x) the element Z11 must be inserted into one of the two clusters.

| PT | x | y |
|---|---|---|
| z1 | -6 | 0 |
| z2 | -4 | 2 |
| z3 | -4 | 0 |
| z4 | -4 | -2 |
| z5 | -2 | 0 |
| z11 | 0 | 0 |
| z6 | 2 | 0 |
| z7 | 4 | 2 |
| z8 | 4 | 0 |
| z9 | 4 | -2 |
| z10 | 6 | 0 |

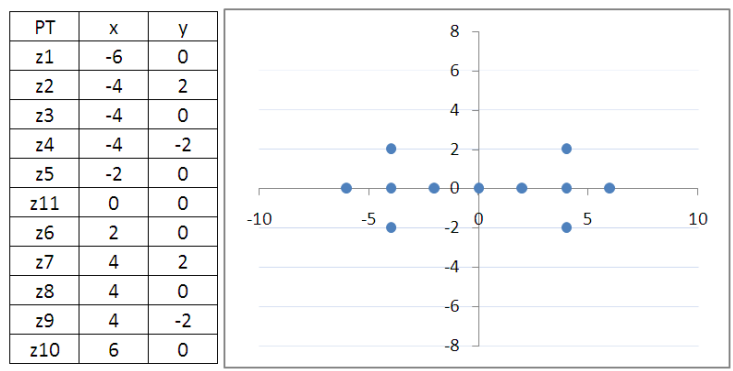

*Fig. 3*

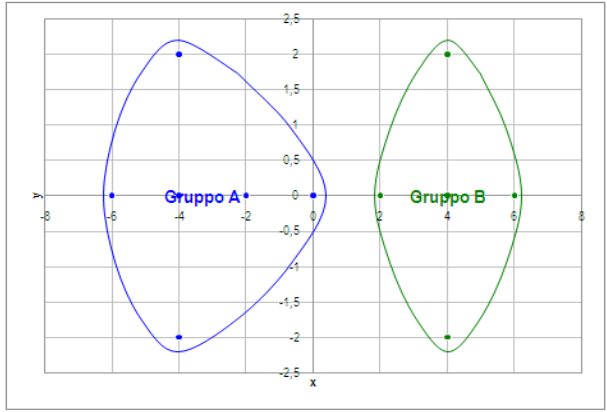

*Fig. 4*

$$U = \begin{bmatrix} 1 & 1 & 1 & 1 & 1 & 1 & 0 & 0 & 0 & 0 & 0 \\ 0 & 0 & 0 & 0 & 0 & 0 & 1 & 1 & 1 & 1 & 1 \end{bmatrix}$$

*tab.2*

Two alternatives are possible. The first is to decide to create three clusters

$$U = \begin{bmatrix} 1 & 1 & 1 & 1 & 1 & 0 & 0 & 0 & 0 & 0 & 0 \\ 0 & 0 & 0 & 0 & 0 & 1 & 0 & 0 & 0 & 0 & 0 \\ 0 & 0 & 0 & 0 & 0 & 0 & 1 & 1 & 1 & 1 & 1 \end{bmatrix}$$

*tab.3*

2. FUZZY CLUSTERING.

The other alternative is to consider a fuzzy clustering, in which the allocation to clusters is not unique, but is represented by the amount of the membership of elements to all clusters calculated. In fuzzy clustering the membership value of each element is not only 0 or 1, but any value that lies between these extremes. The conditions are:

$$\mu_{ik} \in [0,1], \quad 1 \leq i \leq c, \quad 1 \leq k \leq N \qquad (3.1a)$$

$$\sum_{i=1}^{c} \mu_{ik} = 1, \quad 1 \leq k \leq N \qquad (3.1b)$$

$$0 < \sum_{k=1}^{N} \mu_{ik} < N, \quad 1 \leq i \leq c \qquad (3.1c)$$

With these conditions the amount of allocation of individual elements to a specific cluster, rather than the other, calculated in the cartesian coordinates of the centroids shown in Fig. 5, appears to be

$$U = \begin{bmatrix} 0{,}78 & 0{,}78 & 0{,}92 & 0{,}78 & 0{,}80 & 0{,}50 & 0{,}20 & 0{,}22 & 0{,}08 & 0{,}22 & 0{,}22 \\ 0{,}22 & 0{,}22 & 0{,}08 & 0{,}22 & 0{,}20 & 0{,}50 & 0{,}80 & 0{,}78 & 0{,}92 & 0{,}78 & 0{,}78 \end{bmatrix}$$

*tab.4*

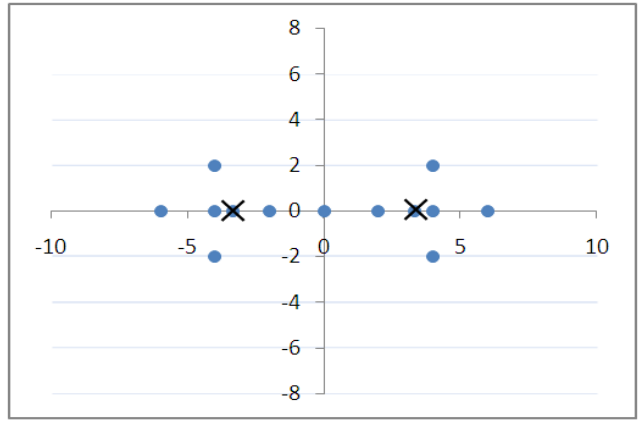

*Fig. 5*

$$cluster1 \ \{-3{,}33; 0\}$$
$$cluster2 \ \ \{3{,}33; 0\}$$

The grades of membership are calculated by reference to the Euclidean distances from the centroid, as determined by the formula

$$d(i, i') = \sqrt{(x_i - x_{i'})^2 + (y_i - y_{i'})^2} \qquad (3.2)$$

|  | elements | | cluster1 centroid | | cluster2 centroid | | crisp clustering | | fuzzy clustering | |
|---|---|---|---|---|---|---|---|---|---|---|
|  | xi1 (x) | xi2 (y) | xi'1 (x) | xi'2 (y) | xi'1 (x) | xi'2 (y) | cd | cd' | fd | fd' |
| z1 | -6,00 | 0,00 | -3,33 | 0,00 | 3,33 | 0,00 | 2,670 | 9,330 | 0,78 | 0,22 |
| z2 | -4,00 | 2,00 | -3,33 | 0,00 | 3,33 | 0,00 | 2,109 | 7,598 | 0,78 | 0,22 |
| z3 | -4,00 | 0,00 | -3,33 | 0,00 | 3,33 | 0,00 | 0,670 | 7,330 | 0,92 | 0,08 |
| z4 | -4,00 | -2,00 | -3,33 | 0,00 | 3,33 | 0,00 | 2,109 | 7,598 | 0,78 | 0,22 |
| z5 | -2,00 | 0,00 | -3,33 | 0,00 | 3,33 | 0,00 | 1,330 | 5,330 | 0,80 | 0,20 |
| z11 | 0,00 | 0,00 | -3,33 | 0,00 | 3,33 | 0,00 | 3,330 | 3,330 | 0,50 | 0,50 |
| z6 | 2,00 | 0,00 | -3,33 | 0,00 | 3,33 | 0,00 | 5,330 | 1,330 | 0,20 | 0,80 |
| z7 | 4,00 | 2,00 | -3,33 | 0,00 | 3,33 | 0,00 | 7,598 | 2,109 | 0,22 | 0,78 |
| z8 | 4,00 | 0,00 | -3,33 | 0,00 | 3,33 | 0,00 | 7,330 | 0,670 | 0,08 | 0,92 |
| z9 | 4,00 | -2,00 | -3,33 | 0,00 | 3,33 | 0,00 | 7,598 | 2,109 | 0,22 | 0,78 |
| z10 | 6,00 | 0,00 | -3,33 | 0,00 | 3,33 | 0,00 | 9,330 | 2,670 | 0,22 | 0,78 |

*Tab.5*

The Z11 element has a factor of 0.5 degree of membership for each cluster. Through fuzzy clustering is no longer required, in extreme cases, to decide in which clusters insert the symmetrical element, because the same element is symmetrical and is exactly half (geometrically speaking) between the two centroids, and the table shows it in an absolutely way. The same table also shows, for example, that the element z3 belongs to cluster 1 significantly more than the elements z1, z2 and z4. At this point it may seem easy to say that these considerations are also largely determined by visual approach to the chart (not much about the correct positioning of the centroids), but try to imagine a much larger and asymmetrical dataset, ie with a much less regular distribution: if there is no evidence of graphic centroid (very probable) the reading of a table like the one described above would be the only source of information for analysis.

3. OUTLIERS PRESENCE.

Let us see the presence of an outlier. For simplicity of exposition we also consider this element symmetrical.

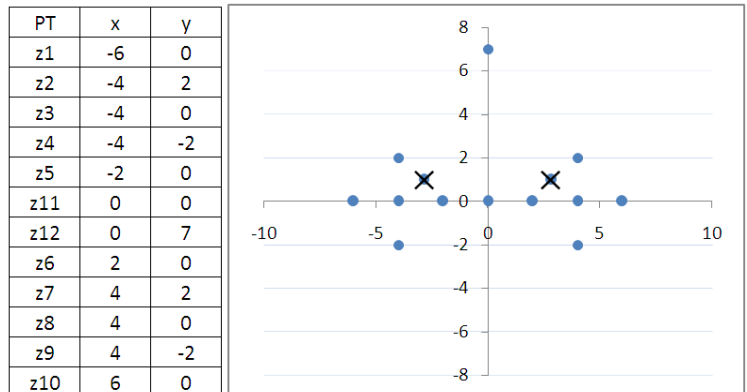

| PT | x | y |
|---|---|---|
| z1 | -6 | 0 |
| z2 | -4 | 2 |
| z3 | -4 | 0 |
| z4 | -4 | -2 |
| z5 | -2 | 0 |
| z11 | 0 | 0 |
| z12 | 0 | 7 |
| z6 | 2 | 0 |
| z7 | 4 | 2 |
| z8 | 4 | 0 |
| z9 | 4 | -2 |
| z10 | 6 | 0 |

*Fig.6*

The position of centroids change, of course

$$cluster1 \ \{-2,85714; 1\}$$
$$cluster2 \ \ \{2,85714; 1\}$$

and with it also changes the weights of the memeberships of the elements to clusters, but not the Z11 element which remains equidistant from the two centroids. And the same happens for Z12 which is also symmetrical by construction.

$$U = \begin{bmatrix} \cdots & 0,50 & 0,50 & \cdots \\ \cdots & 0,50 & 0,50 & \cdots \end{bmatrix}$$

*tab.6*

Here the tables of fuzzy memberships:

$$U = \begin{bmatrix} 0,73 & 0,82 & 0,82 & 0,70 & 0,79 & 0,50 & 0,50 & 0,21 & 0,30 & 0,18 & 0,18 & 0,27 \\ 0,27 & 0,18 & 0,18 & 0,30 & 0,21 & 0,50 & 0,50 & 0,79 & 0,70 & 0,82 & 0,82 & 0,73 \end{bmatrix}$$

*tab.7*

|  | elements | | cluster1 centroid | | cluster2 centroid | | crisp clustering | | fuzzy clustering | |
|---|---|---|---|---|---|---|---|---|---|---|
|  | xi1 (x) | xi2 (y) | xi'1 (x) | xi'2 (y) | xi'1 (x) | xi'2 (y) | cd | cd' | fd | fd' |
| z1 | -6,00 | 0,00 | -2,86 | 1,00 | 2,86 | 1,00 | 3,298 | 8,913 | 0,73 | 0,27 |
| z2 | -4,00 | 2,00 | -2,86 | 1,00 | 2,86 | 1,00 | 1,519 | 6,930 | 0,82 | 0,18 |
| z3 | -4,00 | 0,00 | -2,86 | 1,00 | 2,86 | 1,00 | 1,519 | 6,930 | 0,82 | 0,18 |
| z4 | -4,00 | -2,00 | -2,86 | 1,00 | 2,86 | 1,00 | 3,210 | 7,485 | 0,70 | 0,30 |
| z5 | -2,00 | 0,00 | -2,86 | 1,00 | 2,86 | 1,00 | 1,317 | 4,959 | 0,79 | 0,21 |
| z11 | 0,00 | 0,00 | -2,86 | 1,00 | 2,86 | 1,00 | 3,027 | 3,027 | 0,50 | 0,50 |
| z12 | 0,00 | 7,00 | -2,86 | 1,00 | 2,86 | 1,00 | 6,646 | 6,646 | 0,50 | 0,50 |
| z6 | 2,00 | 0,00 | -2,86 | 1,00 | 2,86 | 1,00 | 4,959 | 1,317 | 0,21 | 0,79 |
| z7 | 4,00 | 2,00 | -2,86 | 1,00 | 2,86 | 1,00 | 6,930 | 1,519 | 0,18 | 0,82 |
| z8 | 4,00 | 0,00 | -2,86 | 1,00 | 2,86 | 1,00 | 6,930 | 1,519 | 0,18 | 0,82 |
| z9 | 4,00 | -2,00 | -2,86 | 1,00 | 2,86 | 1,00 | 7,485 | 3,210 | 0,30 | 0,70 |
| z10 | 6,00 | 0,00 | -2,86 | 1,00 | 2,86 | 1,00 | 8,913 | 3,298 | 0,27 | 0,73 |

*Tab.8*

What said now is technically and mathematically correct. Having the need to clustering dataset into subsets, point of interest is just the centroid, or geometric center, the average distances. And as said in the beginning, the identification of one or more outliers is particulary significative.
The table does not perceive the fact that Z12 is an outlier (with respect to the Z11 that has the same membership values).

4. POSSIBILISTIC CLUSTERING.

A more general form of fuzzy partition is the possibilistic partition.

You get that reducing the constraint (3.1b), while not eliminate it, and then keeping to the memebership value of the item to a subset at least greater than zero.

The conditions thus become:

$$\mu_{ik} \in [0,1], \ 1 \leq i \leq c, \ 1 \leq k \leq N \quad (5.1a)$$

$$\exists i, \mu_{ik} > 0, \forall k \quad (5.1b)$$

$$0 < \sum_{k=1}^{N} \mu_{ik} < N, \ 1 \leq i \leq c \quad (5.1c)$$

The result does not necessarily provide membership value to clusters whose sum is equal to 1, but could be lower if the item is less typical than others. All values that have a sum of the attribution <1 values can be considered as atypical: not strictly outliers in the context of the global dataset, as possible outliers against clusters obtained after reiterations.

The literature in this regard has proposed since the late nineties a series of algorithms for fuzzy clustering (FCM), possibilistic clustering (PCM), the combination of two (FPCM), and others methods built in according to the specific need to study (Modified Possibilistic Fuzzy c-means MFPCM, Suppressed fuzzy c-means Modified MS-FCM, Relational Fuzzy c-means RFCM, Non-Euclidean Relational Fuzzy c-means NERFCM).

In several of these cases the Euclidean distance that has been used in the examples above described is no longer used. Other algorithms for calculating the distance of the elements are often preferred: squared Euclidean distance, city-block (Manhattan) distance, Chebychev distance, power distance, Percent disagreement, Mahalanobis distance. This last has long been considered one of the measures capable in outliers identification, even though it suffered further studies and modifications to make it suitable for various practical cases.

Since the detailed study of all these criteria partition is not the purpose of this document, we will only provide arguments on how to distinguish a possible outlier. The reader will have opportunity to read documents listed in the bibliography, in case he needs detailed statistical arguments.

A good system for the identification of an outlier from a fuzzy system is to use one of the measures listed above: the squared Euclidean distance.

$$d^2(i, i') = (x_i - x_{i'})^2 + (y_i - y_{i'})^2 \qquad (5.2)$$

In this way the weight of the elements becomes progressively greater, and greater distances are emphasized.

The table below shows the results of processing.

|     | elements |        | cluster1 centroid |         | cluster2 centroid |         | crisp clustering |        | fuzzy clustering |      |
|-----|----------|--------|-------------------|---------|-------------------|---------|------------------|--------|------------------|------|
|     | xi1 (x)  | xi2 (y)| xi'1 (x)          | xi'2 (y)| xi'1 (x)          | xi'2 (y)| cd               | cd'    | fd               | fd'  |
| z1  | -6       | 0      | -2,86             | 1,00    | 2,86              | 1,00    | 10,878           | 79,449 | 0,88             | 0,12 |
| z2  | -4       | 2      | -2,86             | 1,00    | 2,86              | 1,00    | 2,306            | 48,020 | 0,95             | 0,05 |
| z3  | -4       | 0      | -2,86             | 1,00    | 2,86              | 1,00    | 2,306            | 48,020 | 0,95             | 0,05 |
| z4  | -4       | -2     | -2,86             | 1,00    | 2,86              | 1,00    | 10,306           | 56,020 | 0,84             | 0,16 |
| z5  | -2       | 0      | -2,86             | 1,00    | 2,86              | 1,00    | 1,735            | 24,592 | 0,93             | 0,07 |
| z11 | 0        | 0      | -2,86             | 1,00    | 2,86              | 1,00    | 9,163            | 9,163  | 0,50             | 0,50 |
| z12 | 0        | 7      | -2,86             | 1,00    | 2,86              | 1,00    | 44,163           | 44,163 | 0,50             | 0,50 |
| z6  | 2        | 0      | -2,86             | 1,00    | 2,86              | 1,00    | 24,592           | 1,735  | 0,07             | 0,93 |
| z7  | 4        | 2      | -2,86             | 1,00    | 2,86              | 1,00    | 48,020           | 2,306  | 0,05             | 0,95 |
| z8  | 4        | 0      | -2,86             | 1,00    | 2,86              | 1,00    | 48,020           | 2,306  | 0,05             | 0,95 |
| z9  | 4        | -2     | -2,86             | 1,00    | 2,86              | 1,00    | 56,020           | 10,306 | 0,16             | 0,84 |
| z10 | 6        | 0      | -2,86             | 1,00    | 2,86              | 1,00    | 79,449           | 10,878 | 0,12             | 0,88 |

*Tab. 9*

You can read the squared Euclidean distances within the columns cd and cd'. In columns fd and fd' the values of fuzzy attributions changes significantly after an updating processing. It is even more clearly possible to distinguish the two clusters: the first formed by the elements z1 … z5 and the other formed by the elements z6 ... z10.

Remain very far the two elements Z11 and Z12, which in this case havs equal attribution levels. But for this type of elements, and only for this, note that the quadratic distance is 9.1 for Z11 and 44.1 for Z12: the distance of this second element is nearly five times higher.

| fuzzy clustering | | | | | | possibilistic fuzzy clust.ing | |
|---|---|---|---|---|---|---|---|
| fd | fd' | d min | clust d max | | | fd | fd' |
| 0,88 | 0,12 | 10,878 | 10,878 | | | 0,88 | 0,12 |
| 0,95 | 0,05 | 2,306 | | | | 0,95 | 0,05 |
| 0,95 | 0,05 | 2,306 | | | | 0,95 | 0,05 |
| 0,84 | 0,16 | 10,306 | | | | 0,84 | 0,16 |
| 0,93 | 0,07 | 1,735 | | | | 0,93 | 0,07 |
| 0,50 | 0,50 | 9,163 | | 1 | 0,5 | 0,50 | 0,50 |
| 0,50 | 0,50 | 44,163 | | 0,207486 | 0,103743 | 0,10 | 0,10 |
| 0,07 | 0,93 | 1,735 | 10,878 | | | 0,07 | 0,93 |
| 0,05 | 0,95 | 2,306 | | | | 0,05 | 0,95 |
| 0,05 | 0,95 | 2,306 | | | | 0,05 | 0,95 |
| 0,16 | 0,84 | 10,306 | | | | 0,16 | 0,84 |
| 0,12 | 0,88 | 10,878 | | | | 0,12 | 0,88 |

*Tab. 10*

At this point, highlighting the distances of the groups using the fuzzy degrees of attribution.

Then find the maximum distance

$$clust\_d_{max} = MAX[d_{min}] \qquad (5.3)$$

An ambiguous element, such as Z11, if it has a distance not exceeding

$$W_{sup} = Q_3 + 1{,}5IQR \qquad (5.4)$$

(where $Q_3$ is the third quartile series distances of the cluster, and IQR is the interquartile range)

$$IQR = Q_3 - Q_1 \qquad (5.5)$$

should not be considered an outlier. Since in our case Z12 is an outlier, the membership degree to clusters while remaining equal, from a point of view possibilistic should highlight the fact that its location makes it less typical than Z11. This can be achieved whereas the distance increases the degree and the memebrship degree should decrease.

Taking as reference the element Z11 distance, although in an atypical position (equal membership values) is not an outlier, and setting the distance in the denominator of an atypical relationship between the two distances, it's possible to recalculates the membership value.

$$fd = \left(\frac{d_{tipic}}{d_{atipic}}\right)/2 \qquad (5.6)$$

The values of the tab. 7 become so

$$U = \begin{bmatrix} 0{,}73 & 0{,}82 & 0{,}82 & 0{,}70 & 0{,}79 & 0{,}50 & 0{,}10 & 0{,}21 & 0{,}30 & 0{,}18 & 0{,}18 & 0{,}27 \\ 0{,}27 & 0{,}18 & 0{,}18 & 0{,}30 & 0{,}21 & 0{,}50 & 0{,}10 & 0{,}79 & 0{,}70 & 0{,}82 & 0{,}82 & 0{,}73 \end{bmatrix}$$

*tab.8*

and are subject to the conditions (5.1x): the membership value is not 1 or 0 as the crisp partition, but it is between 1 and 0 as the fuzzy partition, and in particular sum of memberships even greater than 0 must be no longer equal to 1 if the item is not typical compared to clusters. In this case a sum of the attribution of 0.20 denotes a highly atypical element.

References.

Advanced Computational Intelligence and Intelligent Informatics, 2009;

F. Hoppner, F. Klawonn, *Obtainig interpretable fuzzy models from fuzzy clustering and fuzzy regression*, IEEE, 2000;

F. Klawonn, F. Hoppner, *What is fuzzy about fuzzy clustering? Understanding and improving the concept of the fuzzifier*, Department of Computer Sciences University of Applied Sciences Braunschweing/Wolfenbuettel.

______________________________